\documentclass[12pt]{article}
\usepackage{amsmath,amssymb}
\usepackage[dvips]{graphicx}

 \numberwithin{equation}{section}
\begin{document}
%
\newcommand{\op}[1]{\hat{#1}} 
\newcommand{\unit}[1]{\,\mathrm{#1}} 
\newcommand{\Real}{\mathbb{R}}
\newcommand{\Complex}{\mathbb{C}}
\newcommand{\Natural}{\mathbb{N}}
\newcommand{\unitelement}{\boldsymbol{1}}
\newcommand{\id}{\mathrm{id}}
\newcommand{\isomorphic}{\cong}
\newcommand{\group}[1]{{#1}} 
\newcommand{\algebra}[1]{\mathfrak{#1}} 
\newcommand{\uea}{\mathcal{U}} 
\newcommand{\Lie}{\mathcal{G}} 
\newcommand{\Hopf}{\mathcal{H}} 
\newcommand{\twist}{\mathcal{F}} 
\newcommand{\covariantTwist}{\mathcal{T}} 
\newcommand{\Poincare}{\mathcal{P}} 
\newcommand{\Lorentz}{\group{SO(1,3)}}
\newcommand{\translations}{\mathcal{T}_4}
\newcommand{\Minkowski}{\mathcal{M}_4}
\newcommand{\RiemannCartan}{\mathcal{U}_4}
\newcommand{\ud}{\mathrm{d}} 
\newcommand{\residual}[1]{\mathcal{O}\left(#1\right)} 
\newcommand{\diag}{\mathrm{diag}} 
\vskip1cm
\vskip1cm
\begin{center}
{\Large\bf Gauging the twisted Poincar\'e symmetry\\\vskip0.3cm as
noncommutative theory of gravitation}

\vskip .7cm

{\bf{\large{M. Chaichian$^a$, M. Oksanen$^a$, A. Tureanu$^a$ and G. Zet$^b$}}\\

\vskip .7cm

{\it $^a$Department of Physics, University of Helsinki\\and Helsinki
Institute of Physics, P.O. Box 64, FIN-00014 Helsinki, Finland\\
$^b$Department of Physics, "Gh. Asachi" Technical University,\\Bd.
D. Mangeron 67, 700050 Iasi, Romania}}

\end{center}

\vskip1cm

\begin{abstract}
Einstein's Theory of General Relativity was formulated as a gauge
theory of Lorentz symmetry by Utiyama in 1956, while the
Einstein-Cartan gravitational theory was formulated by Kibble in
1961 as the gauge theory of Poincar\'e transformations. In a
noncommutative space-time with canonical commutation relations
between the coordintes, Lorentz symmetry is violated and field
theories constructed on such space-times have instead the so-called
twisted Poincar\'e invariance. In this paper a gauge theory
formulation of noncommutative gravity is proposed based on the
twisted Poincar\'e symmetry together with the requirement of
covariance under the general coordinate transformations, an
essential ingredient of the theory of general relativity. The
advantages of such a formulation as well as the related problems are
discussed and possible ways out are outlined.
\end{abstract}


\section{Introduction}
It is generally expected that the smooth manifold structure of the
classical space-time should break down at distances of the order of
the Planck length,
\begin{equation}
\label{eq:Planck-length} l_\mathrm{P} = \sqrt{\frac{\hbar
G}{c^3}}\approx 1.6\cdot10^{-35} \unit{m}\ ,
\end{equation}
so that all physical phenomena become essentially \emph{nonlocal}
--- as opposed to the locality of traditional geometrical theories
of gravitation and quantum and gauge field theories of particle
physics. It is hoped that an appropriate implementation of the
nonlocality will eventually enable the formulation of a unified
theory of the fundamental interactions of Nature, which should be
free from singularities, divergences and any other kind of
inconsistences. The noncommutativity of space-time coordinates is
one way to implement the nonlocality of Planck scale physics, which
is well motivated.

Formally, the noncommutativity of coordinate operators $x^\mu$,
$\mu=0,1,2,3$, is achieved by imposing the commutation relations
\begin{equation}
\label{eq:coordinate-noncommutativity} \bigl[ \op{x}^\mu, \op{x}^\nu
\bigr] = i\theta^{\mu\nu}\ ,
\end{equation}
where in the canonical case $\theta^{\mu\nu}$ is an antisymmetric
constant matrix of dimension length-squared, and by letting the
fields on noncommutative space-time be functions of the
noncommutative coordinate operators. Through Weyl quantization the
noncommutative algebra of operators generated by
\eqref{eq:coordinate-noncommutativity} can be represented on the
algebra of ordinary functions on classical space-time by using the
noncommutative Moyal $\star$-product. The more general case with
$\theta^{\mu\nu}$ being an antisymmetric tensor field has also been
considered.

The idea that space-time coordinates do not commute can be seen as a
generalization of the corresponding property of the quantum
mechanical phase space of coordinate $\op{x}^i$ and momentum
$\op{p}_j$ operators,
\begin{equation}
\label{eq:heisenberg-commutation-relation} [\op{x}^i, \op{p}_j] =
i\hbar\delta^i_j \ .
\end{equation}
The first ``quantized space-time'', which was based on a
noncommutative algebra of coordinate operators, was introduced in
\cite{snyder:1947:1}.

Combining Einstein's theory of general relativity and quantum
mechanical measurements obeying Heisenberg's uncertainty principle
leads to operational noncommutativity of space-time coordinates
\cite{doplicher+fredenhagen+roberts:1994,doplicher+fredenhagen+roberts:1995}.
It is not possible to measure distances of the order of the Planck
length, because it would require so much energy to be present in the
localization region of a measurement, that the gravitational
interaction would prevent any signal from leaving the region, i.e.
the localization of the measurement would collapse under its own
gravity. This has led to the formulation of QFT on noncommutative
space-time.

String theory is one of the strongest motivations for considering
noncommutative space-time geometries and noncommutative gravitation.
It has been shown that when the end points of strings in a theory of
open strings are constrained to move on $D$-branes in a constant
(supergravity) $B$-field background and the theory is taken in a
certain low-energy limit, then the full dynamics of the theory is
described by a gauge theory on a noncommutative space-time
\cite{seiberg+witten:1999}. In this Seiberg-Witten (low-energy)
limit \cite{seiberg+witten:1999}, the open string modes completely
decouple from the closed string modes and only the end point degrees
of freedom for the open strings are left to live on a noncommutative
space-time defined by the coordinate commutation relations
\eqref{eq:coordinate-noncommutativity}. Thus noncommutative gauge
theory emerges as a low-energy limit of open string theory with
constant antisymmetric background field.

The formulation of local (gauge) symmetries on a noncommutative
(nonlocal) space-time is a delicate issue. Most gauge groups can not
be defined on noncommutative space-time, because they do not close
under the $\star$-product. The noncommutative unitary group
$U_\star(n)$ can be defined. Its representations, however, are
limited by the no-go theorem
\cite{chaichian+presnajder+sheikh-jabbarii+tureanu:2002} (see also
\cite{terashima:2000,gracia-bondia+martin:2000}) stating that only
the fundamental, anti-fundamental and adjoint representations of the
gauge group $U_\star(n)$ are allowed and matter fields can be
charged under at most two noncommutative simple gauge groups. A
noncommutative Standard Model based on the gauge groups $U_\star(n)$
has been constructed \cite{NCSM} (see also, for its extension to
noncommutative MSSM, \cite{NCMSSM}).

Another approach to the noncommutative gauge theories has been
through the so-called Seiberg-Witten map \cite{
seiberg+witten:1999}, which originally related a noncommutative
$U_\star(n)$ gauge theory to a commutative one, obtained as
low-energy effective limits in string theory, by using two different
regularization methods (the point-splitting method and the
Pauli-Villars method, respectively). The philosophy behind the
Seiberg-Witten map has been subsequently used to extend the possible
noncommutative gauge groups to include special unitary groups as
well: indeed, noncommutative gauge theories with gauge fields valued
in the enveloping algebra of $su(n)$ have been constructed
\cite{Madore,jurco+schraml+schupp+wess:2000} and a corresponding
noncommutative version of the Standard Model has been built
\cite{NCSM-Wess}.

A new interpretation of the relativistic invariance of  the
commutation relations \eqref{eq:coordinate-noncommutativity} was
proposed in \cite{chaichian+kulish+nishijima+tureanu:2004}: by using
the concept of twisted Poincar\'e algebra, the relativistic
invariance can be generalized to the framework of Hopf algebras. If
in the usual (commutative) case, relativistic invariance means
invariance under the Poincar\'e transformation, then in the
noncommutative case relativistic invariance means invariance under
twisted Poincar\'e transformations \cite{
chaichian+presnajder+tureanu:2005}. Since the twist deformation of
the Poincar\'e algebra does not affect the multiplication of the
algebra generators, the structure of the algebra is preserved, and
consequently the representation content of the twisted Poincar\'e
algebra is identical with that of the usual Poincar\'e algebra. This
legitimates the usage of the familiar representations of the
Poincar\'e symmetry in the context of noncommutative field theories
\cite{ chaichian+kulish+nishijima+tureanu:2004}. The noncommutative
field theories, although they lack the Lorentz symmetry, are
invariant under the twisted Poincar\'e algebra, deformed by the
Abelian twist element
\begin{equation}
\label{eq:abelian-twist-element} \twist =
e^{\frac{i}{2}\theta^{\mu\nu}P_\mu\otimes P_\nu} \ ,
\end{equation}
where $P_\mu=-i\partial_\mu$ are the generators of space-time
translations. The twist induces, on the representations of the
Poincar\'e algebra, the deformed multiplication
\begin{equation}
\label{eq:star-multiplication} \mu(\phi\otimes\psi) = \phi\psi
\longrightarrow \mu_\star(\phi\otimes\psi) =
\mu\bigl(\twist^{-1}(\phi\otimes\psi)\bigr) = \phi\star\psi \,,
\end{equation}
which is precisely the (Moyal) $\star$-product. The question about
the action of the twisted Poincar\'e algebra on fields and of the
actual meaning of the invariance under twisted Poincar\'e algebra
has been raised in \cite{
chaichian+kulish+tureanu+zhang+zhang:2007}, where it was proposed to
seek the answer by re-constructing the fields using the method of
induced representations, in a manner compatible with the twisted
Poincar\'e algebra. Along these lines of thought, a new
interpretation of the noncommutative fields has been proposed in
\cite{chaichian+nishijima+salminen+tureanu:2008}, according to which
the noncommutative fields carry representations of the full Lorentz
group, but admit transformations only under the residual symmetry
which preserves the matrix $\theta_{\mu\nu}$ invariant.

Recently, an attempt was made to twist also the gauge algebra, by
extending the global Poincar\'e algebra through a semidirect product
with the gauge algebra, and by twisting the coproduct of the
combined algebra by using the Abelian twist element
\eqref{eq:abelian-twist-element}. This approach was shown
\cite{chaichian+tureanu:2006} to be in conflict with the very idea
of gauge symmetry, since it implicitly assumed that when a field
transforms according to a given representation, then its partial
derivatives of any order also transform in the same representation
of the gauge algebra, which is obviously not the case.

The question arises whether the concept of twist provides a
\emph{symmetry principle} for formulating noncommutative field
theories: meaning that any symmetry such theories may enjoy, be it
space-time or internal, global or local, should be formulated as a
twisted symmetry. It has been shown
\cite{chaichian+tureanu+zet:2007} that it is not possible to twist
the internal gauge transformations and at the same time keep the
Moyal space-time structure defined by
\eqref{eq:coordinate-noncommutativity}. It is intriguing that the
\emph{external} Poincar\'e symmetry and the \emph{internal} gauge
symmetry can not be unified under a common twist. This situation is
reminiscent of the Coleman-Mandula theorem
\cite{coleman+mandula:1967}, although not entirely, since this
theorem concerns global symmetry and simple groups. However one can
envisage that supersymmetry \cite{haag+lopuszansky+sohnius:1975},
due to its intrinsic internal symmetry, may reverse the situation,
and a noncommutative gauge theory may be constructed by means of a
twist.

There is a good understanding of the noncommutative effects on
matter and gauge fields defined on the flat noncommutative
space-time. The next step is to incorporate gravity by considering
curved noncommutative space-times. The main problem is that the
noncommutativity parameter $\theta^{\mu\nu}$ is usually taken to be
constant, which breaks the Lorentz invariance of the commutation
relations \eqref{eq:coordinate-noncommutativity}, and implicitly of
any noncommutative field theory. This has motivated a large amount
of work to study noncommutative deformations of general relativity
(see, e.g.,
\cite{chamseddine:2001:1}--\cite{chaichian+tureanu+setare+zet:2008}
and references therein). Noncommutative gauge theory defined through
matrix models
\cite{steinacker:2007,grosse+steinacker+wohlgenannt:2008} contains a
specific version of gravity as an intrinsic part, and provides a
dynamical theory on noncommutative spaces. Noncommutative
deformations of gravity have also led to a complex metric and gauge
groups larger than the Lorentz group
\cite{chamseddine:2001:1,chamseddine:2003}. A noncommutative General
Relativity restricted to the volume-preserving transformations
(unimodular theory of gravity) has been also constructed
\cite{calmet+kobakhidze:2005}. Lately, the version of noncommutative
gravity obtained by the deformation of the diffeomorphism algebra
\cite{ aschieri+blohmann+dimitrijevic+meyer+schupp+wess:2005} using
the twist introduced in
\cite{chaichian+kulish+nishijima+tureanu:2004} has been most studied
in the literature. However, it turned out that the dynamics of the
noncommutative gravity arising from string theory
\cite{alvarez-gaume+meyer+vasquez-mozo:2006} is much richer than
this version of noncommutative gravity. The dynamics of closed
strings in the presence of a constant $B$-field induces a
gravitational action in the next-to-leading order in the
Seiberg-Witten limit \cite{seiberg+witten:1999}. Some of the
three-graviton vertices have been derived and they can not be
obtained from an action written only in terms of the
$\star$-product. It is suspected that the reason for this is the
non-invariance of the Moyal $\star$-product under space-time
diffemorphisms. A geometrical approach to noncommutative gravity,
leading to a general theory of noncommutative Riemann surfaces in
which the problem of the frame-dependence of the $\star$-product is
also recognized, has been proposed in \cite{CTZZ}.

A possibility to obtain a theory which is covariantly deformed under
the local Poincar\'e transformations is that of gauging the twisted
Poincar\'e algebra itself. Einstein's Theory of General Relativity
was formulated as a gauge theory of Lorentz symmetry by Utiyama
\cite{utiyama:1956} in 1956, while the Einstein-Cartan gravitational
theory was formulated by Kibble \cite{kibble:1961} in 1961, as the
gauge theory of Poincar\'e transformations. Instead of the partial
derivatives in the Abelian twist element
\eqref{eq:abelian-twist-element} one can use the covariant
derivatives \cite{ kibble:1961} (see also
\cite{book:blogojevic:2001}):
\begin{equation}
\nabla_\mu = \partial_\mu + \frac{i}{2}
\omega_\mu^{\phantom{\mu}ab}\Sigma_{ab} \ ,
\end{equation}
where the (constant) matrices $\Sigma_{ab}$ form a representation of
the Lorentz algebra. We can define a covariant non-Abelian twist
element as
\begin{equation}
\label{eq:covariant-twist} \covariantTwist =
e^{-\frac{i}{2}\theta^{\mu\nu} \nabla_\mu \otimes \nabla_\nu +
\residual{\theta^2}} \ ,
\end{equation}
with possible covariant higher order terms in the noncommutativity
parameter  $\theta^{\mu\nu}$ in the exponent. In this paper we study
the properties of such a covariant twist.

In Section~\ref{sec:commutative-gauge-theory-of-gravitation} we
review the commutative gauge theory of gravitation. The gauge
covariant derivative is defined by considering the Poincar\'e
algebra as gauge symmetry. The curvature and torsion tensors are
obtained from the commutator of covariant derivatives. These results
show that the Poincar\'e gauge theory of gravitation has the
structure of a Riemann-Cartan space $\RiemannCartan$ with both
curvature and torsion.

In Section~\ref{sec:twisting-Poincare-algebra}, the mathematical
framework of twisted Hopf algebras is explained and the twisted
Poincar\'e algebra is defined. The concept of relativistic
invariance in the noncommutative theory is understood as invariance
under the twisted Poincar\'e transformations.

Section~\ref{sec:gauging-twisted-Poincare} is devoted to the
possibility of gauging the twisted Poincar\'e symmetry itself. A
covariant non-Abelian twist element is defined by using the
covariant derivative of the Poincar\'e gauge theory. The conditions
ensuring that the Hopf algebra structure is preserved by the twist
are verified. It is shown that the $\star$-product induced by the
covariant twist is not associative. Therefore, the twisted
Poincar\'e symmetry can not be gauged by generalizing the Abelian
twist \eqref{eq:abelian-twist-element} to a covariant non-Abelian
twist \eqref{eq:covariant-twist}, nor by introducing a more general
covariant twist element.

\section{Commutative gauge theory of gravitation}
\label{sec:commutative-gauge-theory-of-gravitation} General
Relativity (GR) still lacks the status of fundamental microscopic
theory, because of the standing problems of quantization of the
gravitational field and the existence of singular solutions under
very general assumptions. Since the concept of gauge symmetry has
been highly successful in describing the other three fundamental
interactions, gauge theories of gravitation are very attractive. The
important role of the Poincar\'e symmetry as the concept of
relativistic invariance in the quantum field theory, leads one to
consider the Poincar\'e gauge symmetry as a natural framework for
describing the gravitational interaction.

In the absence of the gravitational field, the underlying space-time
symmetry is traditionally described by the global Poincar\'e group.
If we want to obtain a physical theory that is invariant under local
Poincar\'e transformations, the parameters of the transformations
will depend on space-time coordinates, and thus new
\emph{compensating} or \emph{gauge} fields have to be introduced.
These fields describe the gravitational interaction and one hopes to
obtain a quantum theory in analogy with the internal gauge theories.
An important point to be emphasized is that the Poincar\'e gauge
theory of gravitation contains GR as a special case. Its geometric
interpretation shows that the space-time has the structure of
Riemann-Cartan geometry, possessing both \emph{curvature} and
\emph{torsion} \cite{kibble:1961}--\cite{hehl+heyde+kerlick:1976}.

The Einstein-Cartan theory of gravitation is a modification of GR,
allowing space-time to have torsion, in addition to curvature, and
relating torsion to the density of intrinsic angular momentum (the
spin). In GR the Lorentz group, instead of the Poincar\'e group, is
the structure group acting on the orthonormal Lorentz frames in the
tangent spaces of the space-time manifold. Therefore, there is no
room for translations in GR and thus for the torsion and spin
tensors. In the Poincar\'e gauge theory, the torsion and its
relation to the spin are naturally introduced, restoring the role of
the Poincar\'e symmetry in relativistic gravity. The curvature and
torsion are surface densities of Lorentz transformations and
translations, respectively.

The Einstein-Cartan theory is a viable theory of gravitation that
differs slightly from Eintein's GR. The effects of spin and torsion
can be significant only when the density of matter is very high, but
nevertheless much smaller than the Planck density at which quantum
gravitational effects are believed  to dominate. It is possible that
the Einstein-Cartan theory will prove to be a better classical limit
for a future quantum theory of gravitation than the theory without
spin.

The (global) Poincar\'e group is a 10-dimensional noncompact Lie
group which has the structure of a semidirect product of the
translation group $\translations$ and of the Lorentz group
$\Lorentz$
\[
\Poincare = \Lorentz \ltimes \translations \ .
\]
In order to define its transformations, we consider the Minkowski
space-time $\Minkowski$, endowed with the real coordinates $x^\mu$,
$\mu=0,1,2,3$. On $\Minkowski$ it is possible to choose global
inertial coordinates, such that the infinitesimal interval has the
form $\ud s^2=\eta_{\mu\nu}\ud x^\mu \ud x^\nu$, where
$\eta_{\mu\nu}=\diag(-1,1,1,1)$ is the Minkowski metric. The
isometry group of $\Minkowski$ is the group of global Poincar\'e
transformations, written in the infinitesimal form as
\begin{equation}
x^{'\mu} = x^\mu + \omega^\mu_{\phantom{\mu}\nu}x^\nu + \epsilon^\mu
\ ,
\end{equation}
where $\omega^{\mu\nu}=-\omega^{\nu\mu}$ and $\epsilon^\mu$ are the
ten infinitesimal parameters associated to the Lorentz rotations and
space-time translations, respectively.

In order to define matter fields on space-time (scalars, vectors,
spinors etc.), we consider the tangent space $T_p$ at each point
$p\in\Minkowski$. For $\Minkowski$ every $T_p$ actually coincides
with the manifold $\Minkowski$ itself (i.e. $\Minkowski$ is flat and
invariable). On each tangent space $T_p$ we can use a coordinate
frame (C), consisting of four vectors $e_\mu$ tangent to the
coordinate lines, or a local Lorentz frame (L) of four orthonormal
vectors $e_a(x)$,
\[
e_a(x) \cdot e_b(x) = \eta_{ab} = \diag(-1,1,1,1) \ ,
\]
which are named the tetrad. The Latin indices ($a,b,\ldots$) refer
to the L-frames and the Greek indices refer to the C-frames. To each
L-frame $\{e_a\}$ we can associate local inertial coordinates $x^a$,
$a=0,1,2,3$. If the coordinates are globally inertial, one can
always choose the tetrad to coincide with the C-frame,
$e_a=\delta_a^\mu e_\mu$.

A matter field $\phi(x)$ on space-time is always given in the
L-frame. In general it is a multi-component object which can be
written as a vector-column. The action of a global Poincar\'e
transformation in $T_p$ transforms each L-frame into another
L-frame, inducing an appropriate (infinitesimal) transformation of
the field $\phi(x)$
\begin{subequations}
\label{eq:global-Poincare-transformation}
\begin{align}
x^{'a} &= x^a + \omega^a_{\phantom{a}b}x^b + \epsilon^a \ ,\\
\phi'(x') &= \left(1 - \frac{i}{2}\omega^{ab}\Sigma_{ab}\right)
\phi(x) \ ,
\end{align}
\end{subequations}
where $\Sigma_{ab}$ are the spin-matrices which act on the field
$\phi(x)$ through matrix-multiplication. For example, if $\phi(x)$
is a spin-$\frac{1}{2}$ Dirac field, then
\begin{equation}
\Sigma_{ab} = \frac{i}{4}[\gamma_a, \gamma_b] \ ,
\end{equation}
where $\gamma_a$ ($a=0,1,2,3$) are the Dirac matrices. If $\phi(x)$
is a spin-1 field, then
\begin{equation}
(\Sigma_{ab})^c_{\phantom{c}d} = i(\delta_a^c \eta_{bd} - \delta_b^c
\eta_{ad}) \ .
\end{equation}

Equivalently we can write \eqref{eq:global-Poincare-transformation}
as
\begin{equation}
\label{eq:internal-like-Poincare-transformation} \delta\phi(x)
\equiv \phi'(x) - \phi(x) = -i\left( \frac{1}{2}\omega^{ab}M_{ab} +
\epsilon^a P_a \right) \phi(x) \ ,
\end{equation}
where
\begin{subequations}
\begin{align}
M_{ab} &= i(x_a\partial_b - x_b\partial_a) + \Sigma_{ab} \equiv L_{ab} + \Sigma_{ab} \ ,\\
P_a &= -i\partial_a \ ,
\end{align}
\end{subequations}
are the generators of the global Poincar\'e transformations in the
space of fields. These generators satisfy the Poincar\'e algebra
$\Poincare$:
\begin{subequations}
\label{eq:Poincare-algebra}
\begin{align}
[P_a, P_b] &= 0 \label{eq:translation-algebra}\\
[M_{ab}, P_c] &= -i(\eta_{ac}P_b - \eta_{bc}P_a) \label{eq:MP-commutation-relations}\\
[M_{ab}, M_{cd}] &= -i(\eta_{ac}M_{bd} - \eta_{ad}M_{bc} -
\eta_{bc}M_{ad} + \eta_{bd}M_{ac}) \ .\label{eq:Lorentz-algebra}
\end{align}
\end{subequations}
The spin-matrices $\Sigma_{ab}$ commute with $L_{ab}$ and $P_a$, and
satisfy the same commutation relations as $M_{ab}$
\eqref{eq:Lorentz-algebra} --- the Lorentz algebra.

Now we consider the local Poincar\'e (gauge) group. In order to make
the Lagrangian $L(\phi, \partial_a\phi)$ invariant under the local
Poincar\'e transformations,
\begin{subequations}
\label{eq:local-Poincare-transformation}
\begin{align}
x^{'a} &= x^a + \omega^a_{\phantom{a}b}(x)x^b + \epsilon^a(x) \ ,\\
\phi'(x') &= \left(1 - \frac{i}{2}\omega^{ab}(x)\Sigma_{ab}\right)
\phi(x) \ ,
\end{align}
\end{subequations}
with the parameters $\omega^{ab}(x)$ and $\epsilon^a(x)$ depending
on space-time coordinates, we have to introduce new compensating
fields $e^a_{\phantom{a}\mu}(x)$ and
$\omega_\mu^{\phantom{\mu}ab}(x)=-\omega_\mu^{\phantom{\mu}ba}(x)$,
named tetrads and spin-connections, respectively \cite{cho:1976}.
They enable us to define  the gauge covariant derivative \cite{
utiyama:1956,kibble:1961} (see also \cite{book:blogojevic:2001}),
which in the C-frame is written as
\begin{equation}
\label{eq:C-covariant-derivative} \nabla_\mu \phi = \left(
\partial_\mu + \frac{i}{2} \omega_\mu^{\phantom{\mu}ab}\Sigma_{ab}
\right) \phi \ .
\end{equation}
The gauge fields $e^a_{\phantom{a}\mu}$ have inverses
$e_a^{\phantom{a}\mu}$ which satisfy
\begin{equation}
\label{eq:tetrad-inverse} e^a_{\phantom{a}\mu} e_b^{\phantom{b}\mu}
= \delta^a_b \ ,\quad e^a_{\phantom{a}\mu} e_a^{\phantom{a}\nu} =
\delta_\mu^\nu \ .
\end{equation}
They can be used to transform C-frame indices $\mu,\nu,\ldots$ into
the L-frame indices $a,b,\ldots$, and vice versa. Thus we can define
the covariant derivative with respect to the L-frame by
\begin{equation}
\label{eq:C-to-L-covariant-derivative} \nabla_a \phi =
e_a^{\phantom{a}\mu}\nabla_\mu \phi \ .
\end{equation}
By introducing \eqref{eq:C-covariant-derivative} into
\eqref{eq:C-to-L-covariant-derivative}, we obtain
\begin{equation}
\label{eq:L-covariant-derivative} \nabla_a \phi = i\left(
e_a^{\phantom{a}\mu}P_\mu + \frac{1}{2}
\omega_a^{\phantom{a}bc}\Sigma_{bc} \right) \phi \ .
\end{equation}
The expression
\begin{equation}
\mathcal{A}_a = i \left( e_a^{\phantom{a}\mu}P_\mu + \frac{1}{2}
\omega_a^{\phantom{a}bc}\Sigma_{bc} \right)
\end{equation}
is also considered as a definition of the gauge potentials valued in
the Lie algebra of the Poincar\'e group
\cite{kibble:1961,book:blogojevic:2001}.

The commutator of two covariant derivatives $\nabla_a$ and
$\nabla_b$ can be calculated by using
\eqref{eq:C-to-L-covariant-derivative} and
\eqref{eq:tetrad-inverse}:
\begin{equation}
[\nabla_a, \nabla_b] \phi = \left(
\frac{1}{2}F^{cd}_{\phantom{cd}ab}\Sigma_{cd} -
F^c_{\phantom{c}ab}\nabla_c \right) \phi \ ,
\end{equation}
where
\begin{equation}
F^{cd}_{\phantom{cd}ab} =
F^{cd}_{\phantom{cd}\mu\nu}e_a^{\phantom{c}\mu}e_b^{\phantom{d}\nu}
\ ,\quad F^c_{\phantom{c}ab} =
F^c_{\phantom{c}\mu\nu}e_a^{\phantom{c}\mu}e_b^{\phantom{d}\nu} \ ,
\end{equation}
and
\begin{align}
R^{ab}_{\phantom{ab}\mu\nu}\equiv F^{ab}_{\phantom{ab}\mu\nu} &= \partial_\mu \omega_\nu^{\phantom{\nu}ab} - \partial_\nu \omega_\mu^{\phantom{\mu}ab} + ( \omega_\mu^{\phantom{\mu}ac}\omega_\nu^{\phantom{\nu}db} - \omega_\nu^{\phantom{\nu}ac}\omega_\mu^{\phantom{\mu}db} )\eta_{cd} \ ,\label{eq:curvature}\\
T^a_{\phantom{a}\mu\nu}\equiv F^a_{\phantom{a}\mu\nu} &=
\partial_\mu e^a_{\phantom{a}\nu} -
\partial_\nu e^a_{\phantom{a}\mu} + (
\omega_\mu^{\phantom{\mu}ab}e^c_{\phantom{c}\nu} -
\omega_\nu^{\phantom{\nu}ab}e^c_{\phantom{c}\mu} )\eta_{bc} \
.\label{eq:torsion}
\end{align}
The quantities $R^{ab}_{\phantom{ab}\mu\nu}$ and
$T^a_{\phantom{a}\mu\nu}$ are identified with the components of the
curvature and torsion tensors of the space-time, respectively.
Therefore the Poincar\'e gauge theory of gravitation has the
geometric structure of the Riemann-Cartan space $\RiemannCartan$
with curvature and torsion.

The metric tensor can be defined by using the tetrad gauge fields.
In a C-frame it has the components
\begin{equation}
\label{eq:metric-tensor} g_{\mu\nu}(x) =
\eta_{ab}e^a_{\phantom{a}\mu}e^b_{\phantom{b}\nu} \ .
\end{equation}
According to \eqref{eq:metric-tensor} the metric itself can be seen
as an effective gauge field, i.e. a dynamical variable.

By imposing the condition of null torsion,
$T^a_{\phantom{a}\mu\nu}=0$, one can solve for the spin-connection
$\omega_\mu^{\phantom{\mu}ab}$ in terms of the tetrads
$e^a_{\phantom{a}\mu}$, thus reducing the Einstein-Cartan theory to
GR.

\section{Twisting the Poincar\'e algebra}
\label{sec:twisting-Poincare-algebra} Space-time noncommutativity is
a way to deform the classical space-time, so that nonlocality
becomes its characteristic feature. This means that the notion of a
point is no longer well-defined. On such a noncommutative space-time
physical phenomena are naturally nonlocal. The coordinates
$\op{x}^\mu$ of noncommutative space-time satisfy the commutation
relations
\begin{equation}
\label{eq:coordinate-commutation-relations} \bigl[ \op{x}^\mu,
\op{x}^\nu \bigr] = i\theta^{\mu\nu}\ ,
\end{equation}
where, in the simplest (canonical) case, $\theta^{\mu\nu}$ is an
antisymmetric
 constant matrix of dimension length-squared. This kind of space-time is described
 by using the methods of \emph{noncommutative geometry}.
 It led to an algebraic description of noncommutative space-times --- based entirely on algebraic functions ---
 and it enabled one to define Yang-Mills gauge theories on a large class of noncommutative spaces.
 For quite some time, the physical applications were based on geometric interpretations of the standard model
 and its various fields and coupling constants.

To describe physics on the noncommutative space-time generated by
\eqref{eq:coordinate-commutation-relations}, one replaces the usual
point-wise product of functions, $f(x)$ and $g(x)$, by the
noncommutative Moyal $\star$-product, which can be written
explicitly as
\begin{multline}
\label{eq:star-product}
(f \star g)(x) = f(x) \exp\left( \frac{i}{2} \overleftarrow{\partial}_\mu \theta^{\mu\nu} \overrightarrow{\partial}_\nu \right) g(x) \\
= f(x)g(x) + \sum_{n=1}^\infty \frac{1}{n!}
\left(\frac{i}{2}\right)^n \theta^{\mu_1 \nu_1}\cdots\theta^{\mu_n
\nu_n} \bigl(\partial_{\mu_1}\cdots\partial_{\mu_n}f(x)\bigr)
\bigl(\partial_{\nu_1}\cdots\partial_{\nu_n}g(x)\bigr) \ .
\end{multline}
Particularly, the commutator of field operators, $\op{\phi}(x)$ and
$\op{\psi}(x)$, is represented on the algebra of functions by the
Moyal bracket:
\begin{equation}
\label{eq:moyal-bracket} [\phi(x), \psi(x)]_\star = \phi(x) \star
\psi(x) - \phi(x) \star \psi(x) \ .
\end{equation}

The defining commutation relations of the noncommutative space-time
\eqref{eq:coordinate-commutation-relations} are clearly not
covariant under Lorentz transformations, because the left-hand side
of the relation is a tensor and the right hand-side is a constant.
Thus the noncommutative space-time does not posses Lorentz symmetry.
This could be a serious problem, because the quantum and gauge field
theories of high energy physics are vitally dependent on the
representation content of the Poincar\'e algebra. The solution to
the problems arising from the breaking of the Lorentz symmetry is
the twisted Poincar\'e symmetry
\cite{chaichian+kulish+nishijima+tureanu:2004,chaichian+presnajder+tureanu:2005}.

The adequate mathematical framework for describing noncommutative
gauge theories and the twisted Poincar\'e symmetry is that of the
Hopf algebras. For Lie algebras one starts with their universal
enveloping algebras, which are the most general unital associative
algebras into which they can be embedded.

Let us consider a Lie algebra $\Lie$ generated by $T_i$,
$i=1,2,\ldots,n$ ,
\begin{equation}
\label{eq:Lie-algebra} [T_i, T_j] \equiv T_i T_j - T_j T_i =
if_{ij}^k T_k \ ,
\end{equation}
where the associative product is $T_i T_j$. Its universal enveloping
algebra $\uea(\Lie)$ consists of the polynomials in the generators
$T_i$ modulo the commutation relations \eqref{eq:Lie-algebra}, and
of the unit element $\unitelement$. The basis of the universal
enveloping algebra can be chosen to consists of $\unitelement$ and
of the fully symmetrized products of the generators
\begin{equation}
T_{(i_1}T_{i_2}\cdots T_{i_n)} \ ,\ n\in\Natural \ .
\end{equation}
The universal enveloping algebra $\uea(\Lie)$ can be extended to a
Hopf algebra $\Hopf$. The algebra $\uea(\Lie)$ consists of a vector
space $V$ over the field $\Complex$ and of the multiplication and
unit linear maps
\begin{align}
m&: V \otimes V \rightarrow V \ ,\\
\eta&: \Complex \rightarrow  V \ ,\nonumber
\end{align}
respectively. Explicitly the multiplication is usually written as
\begin{equation}
m(X\otimes Y) = XY \quad;\ X,Y\in\uea(\Lie) \ .
\end{equation}
The multiplication $m$ is associative
\begin{equation}
m\circ(\unitelement\otimes m) = m\circ(m\otimes\unitelement)
\end{equation}
and the unit map $\eta$ implies the existence of a unit element
$\unitelement$ in $V$
\begin{equation}
m\circ(\unitelement\otimes\eta) = m\circ(\eta\otimes\unitelement) =
\id \ \text{(identity map)} \ .
\end{equation}
The bialgebra structure for $\uea(\Lie)$ is constructed by
introducing the \emph{coproduct} and \emph{counit} homomorphisms
\begin{align}
\Delta &: V \rightarrow V \otimes V \ ,\\
\varepsilon &: V \rightarrow \Complex \ ,\nonumber
\end{align}
respectively. The coproduct $\Delta$ is coassociative,
\begin{equation}
(\id\otimes\Delta)\Delta = (\Delta\otimes\id)\Delta \ ,
\end{equation}
and the counit $\varepsilon$ satisfies
\begin{equation}
(\id\otimes\varepsilon)\circ\Delta =
(\varepsilon\otimes\id)\circ\Delta \ .
\end{equation}
The Hopf algebra is completed by introducing the \emph{antipode}
$S$, an antihomomorphism that is compatible with the bialgebra
structure
\begin{equation}
S: V \rightarrow V \ ,\ m\circ(S\otimes\unitelement)\circ\Delta =
m\circ(\unitelement\otimes S)\circ\Delta = \eta\circ\varepsilon \ .
\end{equation}

The Hopf algebra structure of $\uea(\Lie)$ is defined by
\begin{align}
\Delta_0(X) &= X\otimes\unitelement + \unitelement\otimes X \ ,& \Delta_0(\unitelement) &= \unitelement\otimes\unitelement \ ,\label{eq:primitive-coproduct}\\
\varepsilon(X) &= 0 \ ,& \varepsilon(\unitelement) &= 1 \ ,\label{eq:counit}\\
S(X) &= -X \ ,& S(\unitelement) &= \unitelement \label{eq:antipode}\
,
\end{align}
for all $X\in V-\{\unitelement\}$. The Hopf algebra $\uea(\Lie)$ is
noncommutative, but cocommutative due to the symmetry of the
coproduct \eqref{eq:primitive-coproduct}.

We can deform a cocommutative Hopf algebra like $\uea(\Lie)$ to a
non-cocommutative one by introducing a twist element
\[
\twist \in \uea(\Lie) \otimes \uea(\Lie)
\]
and by redefining the coproduct of the Hopf algebra by a similarity
transformation
\begin{equation}
\label{eq:twisted-coproduct} \Delta_0(X) \longrightarrow \Delta_t(X)
= \twist \Delta_0(X) \twist^{-1}\ , \quad X\in \Lie \ ,
\end{equation}
in other words by twisting the coproduct of $\uea(\Lie)$
\cite{Drinfeld} (see also the monographs \cite{monographs}). In
order to preserve the Hopf algebra structure, the twist element has
to satisfy the twist conditions
\begin{align}
\label{eq:first-twist-condition}
\twist_{12} (\Delta_0\otimes\id) \twist &= \twist_{23} (\id\otimes\Delta_0) \twist \ ,\\
\label{eq:last-twist-condition} (\varepsilon\otimes\id) \twist &=
\unitelement = (\id\otimes\varepsilon) \twist \ ,
\end{align}
where $\twist_{12}=\twist\otimes\unitelement$ and
$\twist_{23}=\unitelement\otimes\twist$. We denote the twist
deformed algebra by $\uea_t(\Lie)$. The twist element does not
affect the multiplication $m$ of the algebra $\uea_t(\Lie)$ and
therefore the commutation relations \eqref{eq:Lie-algebra} among the
generators of $\uea(\Lie)$ are preserved. This means that the
representation content of $\uea_t(\Lie)$ is identical with that of
$\uea(\Lie)$. What is affected by the twist, is the action of
$\uea_t(\Lie)$ onto the tensor products of its representations, i.e.
the \emph{Leibniz rule}.

The solution to the problem of representations in noncommutative
quantum field theory, due to the non-invariance under Lorentz
transformations, was proposed in
\cite{chaichian+kulish+nishijima+tureanu:2004,chaichian+presnajder+tureanu:2005}
in the form of the twisted Poincar\'e symmetry. A twist deformation
of the universal enveloping algebra $\uea(\Poincare)$ of the
Poincar\'e algebra $\Poincare$ was introduced, providing a new
symmetry that is respected by the noncommutative theory obtained by
Weyl quantization on the noncommutative space-time
\eqref{eq:coordinate-commutation-relations}. Since the twist
deformation does not alter the multiplication in $\uea(\Poincare)$,
the commutation relations among its generators
\eqref{eq:Poincare-algebra} are preserved. Thus \emph{the
representation content of the twisted algebra $\uea_t(\Poincare)$ is
the same as the representation content of the usual Poincar\'e
algebra}. This legitimates the usage of the familiar representations
of the Poincar\'e symmetry in the context of noncommutative field
theories.

The Poincar\'e algebra $\Poincare$ \eqref{eq:Poincare-algebra} has a
commutative subalgebra of translation generators
$P_\mu=-i\partial_\mu$ that can be used to construct the
\emph{Abelian} twist element
\begin{equation}
\label{eq:abelian-twist} \twist = e^{\frac{i}{2} \theta^{\mu\nu}
P_\mu \otimes P_\nu} \ ,
\end{equation}
where $\theta^{\mu\nu}$ is the real constant antisymmetric matrix in
\eqref{eq:coordinate-commutation-relations}. It clearly satisfies
the twist conditions
\eqref{eq:first-twist-condition}--\eqref{eq:last-twist-condition}
and thus it can be used to consistently twist the coproduct
\eqref{eq:primitive-coproduct} of the Hopf algebra
$\uea(\Poincare)$. The coproduct of the translation generators
$P_\mu$ is not affected by the twist \eqref{eq:abelian-twist} due to
the commutativity of translations \eqref{eq:translation-algebra},
\begin{equation}
\label{eq:coproduct-of-P} \Delta_t(P_\mu) = \Delta_0(P_\mu) =
P_\mu\otimes\unitelement + \unitelement\otimes P_\mu \ .
\end{equation}
The coproduct of the Lorentz generators $M_{\mu\nu}$ is altered by
the twist, because of their non-vanishing commutation relations with
$P_\mu$ (see \eqref{eq:MP-commutation-relations}),
\begin{multline}
\label{eq:coproduct-of-M} \Delta_t(M_{\mu\nu}) =
\Delta_0(M_{\mu\nu}) - \frac{1}{2}\theta^{\rho\sigma} \bigl(
(\eta_{\rho\mu}P_\nu - \eta_{\rho\nu}P_\mu) \otimes P_\sigma 
+ P_\rho \otimes (\eta_{\sigma\mu}P_\nu - \eta_{\sigma\nu}P_\mu)
\bigr) .
\end{multline}

Since the Abelian twist element \eqref{eq:abelian-twist} only
involves the generators $P_\mu$, only the coordinate dependency of
the fields $\phi(x)$ is involved in the deformed multiplication of
the fields. Therefore the matrix-valued generators $\Sigma_{\mu\nu}$
act on the component degrees of freedom of the fields $\phi(x)$ in
the same way, in the deformed and non-deformed algebra cases, i.e.
through the matrix multiplication and the symmetric coproduct
\begin{equation}
\Delta_t(\Sigma_{\mu\nu}) = \Delta_0(\Sigma_{\mu\nu}) =
\unitelement\otimes\Sigma_{\mu\nu} +
\Sigma_{\mu\nu}\otimes\unitelement \ .
\end{equation}
It should, however, be mentioned that the definition of fields on
noncommutative space-time is more involved than in the commutative
theory
\cite{chaichian+kulish+tureanu+zhang+zhang:2007,chaichian+nishijima+salminen+tureanu:2008}.

The noncommutative quantum field theories built through Weyl
quantization and the canonical $\star$-product
\eqref{eq:star-product} possess the twisted Poincar\'e symmetry,
which represents the concept of relativistic invariance in
noncommutative field theories. This also enables us to adopt the
point of view according to which the noncommutativity of coordinates
\eqref{eq:coordinate-commutation-relations} is required by the
twisted Poincar\'e symmetry of space-time.

\section{Gauging the twisted Poincar\'e symmetry}
\label{sec:gauging-twisted-Poincare} The local Poincar\'e symmetry
is an \emph{external} gauge symmetry. Through geometrical
interpretation the Poincar\'e gauge symmetry translates to the
covariance under general coordinate transformations and to the local
Lorentz symmetry. This ``duality'' of the Poincar\'e gauge symmetry
is both a problem and a possibility, since it has been shown that an
internal gauge symmetry can not be twisted together with the
Poincar\'e symmetry
\cite{chaichian+tureanu:2006,chaichian+tureanu+zet:2007}. We can
attempt to gauge the twisted Poincar\'e algebra itself and find out
whether the gauge theory of the Poincar\'e symmetry on
noncommutative space-time can be formulated by means of a
gauge-covariant twist.

We could take the direct naive approach and try to construct a
noncommutative gauge theory of the twisted Poincar\'e symmetry by
using the Abelian twist \eqref{eq:abelian-twist} and by replacing
the point-wise product of functions with the Moyal $\star$-product
in the classical theory constructed in
Section~\ref{sec:commutative-gauge-theory-of-gravitation}. The
result would, however, be an inconsistent frame-dependent theory
(due to the frame-dependence of the $\star$-product)
--- in many ways similar to those already developed --- that can
not be a plausible theory of gravitation. We would not be able to
give any meaningful geometrical interpretation to a theory of this
type.

Since the global Poincar\'e symmetry is twisted with the Abelian
twist \eqref{eq:abelian-twist} in the case of the flat
noncommutative space-time, also the generalized Poincar\'e gauge
symmetry on noncommutative space-time should be a quantum symmetry.
A natural way to generalize the Poincar\'e gauge symmetry to the
noncommutative setting is to consider it as a twisted gauge
symmetry, so that the global twisted Poincar\'e symmetry is obtained
in the limit of vanishing gauge fields. When the global twisted
Poincar\'e symmetry is generalized to a gauge symmetry, we have to
introduce the gauge fields in order to compensate the non-covariance
of the partial derivatives, similarly as in the commutative case.
Partial derivatives $\partial_\mu$ will be replaced by covariant
derivatives, which in the coordinate frame read
\begin{equation}
\label{eq:covariant-derivative-for-twisted-poincare-gauge-theory}
\nabla_\mu = d_\mu + \mathcal{A}_\mu(x) = i\left(
e^a_{\phantom{a}\mu}(x) P_a +
\frac{1}{2}\omega_\mu^{\phantom{\mu}ab}(x) \Sigma_{ab} \right) \ ,
\end{equation}
where the $\Sigma_{\nu\rho}$ generate a finite-dimensional
representation of the Lorentz algebra. The difference compared to
the covariant derivative of an internal gauge symmetry
\cite{chaichian+tureanu+zet:2007}
\begin{equation}
\label{eq:covariant-derivative-for-internal-gauge-theory} D_\mu =
\partial_\mu + iA_\mu(x) = i(P_\mu + A_\mu^a(x) T_a)
\end{equation}
are the tetrad gauge fields $e^a_{\phantom{a}\mu}$ multiplying
$P_\mu$ in
\eqref{eq:covariant-derivative-for-twisted-poincare-gauge-theory}.
$\mathcal A_\mu$ are the gauge fields associated to the local
Lorentz transformations. In order to obtain a theory that is
covariantly deformed under the Poincar\'e gauge transformations, the
frame-dependent translation generators $P_\mu$ have to be replaced
by the covariant derivatives $-i\nabla_\mu$ in the Abelian twist
element \eqref{eq:abelian-twist}. The covariant non-Abelian twist
element is of the form
\begin{equation}
\label{eq:twist-for-poincare-gauge-theory} \covariantTwist =
e^{-\frac{i}{2}\theta^{\mu\nu} \nabla_\mu \otimes \nabla_\nu +
\residual{\theta^2}} \ ,
\end{equation}
where $\residual{\theta^2}$ stands for the possible additional
covariant terms in higher orders of the noncommutativity parameter
$\theta^{\mu\nu}$\footnote{The following discussion is presented for
the exponential form \eqref{eq:twist-for-poincare-gauge-theory}, but
the results are valid for any invertible functional form.}. Because
of the similar forms of the covariant derivatives
\eqref{eq:covariant-derivative-for-twisted-poincare-gauge-theory}
and \eqref{eq:covariant-derivative-for-internal-gauge-theory} and of
their twist elements, the basic algebraic reasoning presented in
\cite{chaichian+tureanu+zet:2007} holds also for the twist element
\eqref{eq:twist-for-poincare-gauge-theory} proposed here. The gauge
fields $\mathcal{A}_\mu$ alone in $\nabla_\mu$ will violate the
twist condition \eqref{eq:first-twist-condition} and the rest of
gauge fields $e^a_{\phantom{a}\mu}$ are not able to rescue the twist
condition. The fact that there are now two second rank (field
strength) tensors \eqref{eq:curvature}--\eqref{eq:torsion} does not
help to satisfy the twist condition.

Following the arguments of \cite{chaichian+tureanu+zet:2007}, we can
attempt to impose the twist condition
\eqref{eq:first-twist-condition}. First we consider the twist
element \eqref{eq:twist-for-poincare-gauge-theory} with only the
first order term in $\theta$ in the exponent. The second order terms
in $\theta$ that do not cancel in the twist condition
\eqref{eq:first-twist-condition} are, in the left-hand side
\begin{align}
\label{eq:lhs-twist-condition}
\frac{1}{2}\left(-\frac{i}{2}\right)^2 \theta^{\mu\nu}\theta^{\rho\sigma} &\bigl( 2\nabla_\mu\nabla_\rho\otimes\nabla_\nu\otimes\nabla_\sigma + 2\nabla_\mu\otimes\nabla_\nu\nabla_\rho\otimes\nabla_\sigma \\
&+ \nabla_\mu\otimes\nabla_\rho\otimes\nabla_\nu\nabla_\sigma +
\nabla_\rho\otimes\nabla_\mu\otimes\nabla_\nu\nabla_\sigma \bigr)
\nonumber
\end{align}
and in the right-hand side
\begin{align}
\label{eq:rhs-twist-condition}
\frac{1}{2}\left(-\frac{i}{2}\right)^2 \theta^{\mu\nu}\theta^{\rho\sigma} &\bigl( 2\nabla_\rho\otimes\nabla_\mu\nabla_\sigma\otimes\nabla_\nu + 2\nabla_\rho\otimes\nabla_\mu\otimes\nabla_\nu\nabla_\sigma \\
&+ \nabla_\mu\nabla_\rho\otimes\nabla_\nu\otimes\nabla_\sigma +
\nabla_\mu\nabla_\rho\otimes\nabla_\sigma\otimes\nabla_\nu \bigr)
\nonumber\ .
\end{align}
These terms can not be canceled by terms that contain second rank
tensors
\begin{equation} \label{eq:poincare-field-strength}
R^{ab}_{\phantom{ab}\mu\nu}\Sigma_{ab} \ ,\quad
T^a_{\phantom{a}\mu\nu}\nabla_a\,,
\end{equation}
because the two indices for such tensors come from the same
$\theta^{\mu\nu}$, unlike for the $\nabla\nabla$ factors in
\eqref{eq:lhs-twist-condition} and \eqref{eq:rhs-twist-condition}.
This is why such terms are not included in twist element
\eqref{eq:twist-for-poincare-gauge-theory} in the first place. The
other possible second order terms in
\eqref{eq:twist-for-poincare-gauge-theory} have the forms
\begin{align}
\theta^{\mu\nu}\theta^{\rho\sigma}\,& 1\otimes\nabla_\mu\nabla_\nu\nabla_\rho\nabla_\sigma \ ,& \theta^{\mu\nu}\theta^{\rho\sigma}\, \nabla_\mu\nabla_\nu\nabla_\rho\nabla_\sigma\otimes1 \ ,\\
\theta^{\mu\nu}\theta^{\rho\sigma}\,& \nabla_\mu\otimes\nabla_\nu\nabla_\rho\nabla_\sigma \ ,& \theta^{\mu\nu}\theta^{\rho\sigma}\, \nabla_\mu\nabla_\nu\nabla_\rho\otimes\nabla_\sigma \ ,\\
\theta^{\mu\nu}\theta^{\rho\sigma}\,&
\nabla_\mu\nabla_\nu\otimes\nabla_\rho\nabla_\sigma \ ,
\end{align}
with all the permutations of indices of the covariant derivatives
--- although the antisymmetry of $\theta$ greatly reduces the number
of independent permutations. We have verified that when introduced
into the twist element \eqref{eq:twist-for-poincare-gauge-theory}
and consequently into the twist condition
\eqref{eq:first-twist-condition}, these second orders terms can
never cancel all the terms in \eqref{eq:lhs-twist-condition} and
\eqref{eq:rhs-twist-condition}. Therefore, the twist condition
\eqref{eq:first-twist-condition} can not be fulfilled in the second
order in $\theta$.

It is well known that the gauging of the translation symmetry leads
to the Einstein-Hilbert Lagrangian and to the covariance under
general coordinate transformations \cite{book:blogojevic:2001}.
Hence, it is interresting to see whether the gauge theory of the
\emph{external} translation symmetry group $\translations$ can be
consistently defined together with the twisted Poincar\'e symmetry.
The covariant derivative for the local translations is
\begin{equation}
\label{eq:covariant-derivative-for-translation-gauge-theory} d_\mu =
i e^a_{\phantom{a}\mu} P_a \ .
\end{equation}
In fact, this is also the covariant derivative of the Poincar\'e
gauge symmetry for one-dimensional representations, for which the
covariant derivative
\eqref{eq:covariant-derivative-for-twisted-poincare-gauge-theory}
should reduce to
\eqref{eq:covariant-derivative-for-translation-gauge-theory}.
Clearly the gauge fields $e^a_{\phantom{a}\mu}$ now contain
contributions also from the local Lorentz transformations. Since the
covariant derivatives of the translation group do not commute,
\begin{equation}
[d_\mu, d_\nu] = C^\rho_{\phantom{\rho}\mu\nu} d_\rho \ ,\quad
C^\rho_{\phantom{\rho}\mu\nu} = (e^a_{\phantom{a}\mu}\partial_a
e^b_{\phantom{b}\nu} - e^a_{\phantom{a}\nu}\partial_a
e^b_{\phantom{b}\mu}) e_b^{\phantom{b}\rho} \ ,
\end{equation}
the covariant element
\begin{equation}
\label{eq:twist-for-translation-gauge-theory} \covariantTwist =
e^{-\frac{i}{2}\theta^{\mu\nu} d_\mu \otimes d_\nu +
\residual{\theta^2}} = e^{\frac{i}{2}\theta^{\mu\nu}
e^a_{\phantom{a}\mu} P_a \otimes e^b_{\phantom{b}\nu} P_b +
\residual{\theta^2}}
\end{equation}
can not be of the Abelian type \eqref{eq:abelian-twist}, which is
known to be a twist. Because of this and the high level of
arbitrariness in choosing the gauge fields $e^a_{\phantom{a}\mu}$ in
the translationally covariant twist
\eqref{eq:twist-for-translation-gauge-theory}, we face similar
algebraic problems as with the covariant twist element
\eqref{eq:twist-for-poincare-gauge-theory} of the full Poincar\'e
gauge symmetry. The twist element
\eqref{eq:twist-for-translation-gauge-theory} does not satisfy the
twist condition \eqref{eq:first-twist-condition}, even though its
form is much simpler now. Thus, it is not only the local Lorentz
symmetry that breaks the validity of the non-Abelian Poincar\'e
gauge covariant twist element
\eqref{eq:twist-for-poincare-gauge-theory}; the external gauge
symmetry associated with the general coordinate transformations is
just as problematic.

Thus, we have obtained the result that the Poincar\'e gauge
covariant non-Abelian element
\eqref{eq:twist-for-poincare-gauge-theory} is not a twist and the
$\star$-product defined by it is not associative. We can conclude
that \emph{the twisted Poincar\'e symmetry can not be gauged by
generalizing the Abelian twist \eqref{eq:abelian-twist} to a
covariant non-Abelian twist
\eqref{eq:twist-for-poincare-gauge-theory}, nor by introducing a
more general covariant twist element}.

It should be mentioned that from the mathematical point of view, we
could try to deform the action of the twisted Poincar\'e algebra on
its representations, instead of generalizing the twist element, but
it seems unlikely that such an approach could solve the problems
related to the frame-dependent twist element
\eqref{eq:abelian-twist}.

\section{Concluding remarks and perspectives}
In this paper we have investigated the possibility of gauging the
twisted Poincar\'e symmetry in order to obtain a noncommutative
gauge theory of gravitation. A covariant non-Abelian twist element
$\covariantTwist$ has been defined by using the covariant derivative
of the commutative Poincar\'e gauge theory. The twist condition that
assures the associativity of the multiplication on the
representations of the twisted Poincar\'e algebra is violated
already in the second order in the noncommutativity parameter
$\theta^{\mu\nu}$. Adding gauge covariant terms of higher orders in
$\theta^{\mu\nu}$ into the definition of the twist $\covariantTwist$
does not improve the result. When we restrict the gauge symmetry to
the translation group $\translations$, we are faced with similar
algebraic problems as in the case of the full Poincar\'e symmetry.
Thus both the local Lorentz symmetry and the local translational
symmetry, associated with the covariance under general coordinate
transformations, violate the twist condition already in the second
order in the parameter $\theta^{\mu\nu}$.

The question of unifying the external (global or local) Poincar\'e
symmetry and the internal gauge symmetry under a common twist
remains an open fundamental problem of noncommutative gauge
theories.

Since the introduction of a gauge covariant twist breaks the
associativity of the algebra of functions on noncommutative
space-time, both in the internal and external  gauge symmetry cases,
we may have to consider space-time geometries that are also
nonassociative, not only noncommutative. Indeed, there exist in the
literature works on constructing nonassociative theories with some
desired properties (see, e.g.,
\cite{nesterov+sabinin:2006,Majid-nonassoc,sasai+sasakura:2006} and
references therein).

The nonassociativity, as well as the noncommutativity, has its
origin in string theory. It is known that in the presence of a
constant background field,  $\omega=B+F$, the noncommutative
geometry is described by the Moyal product which is associative
\cite{seiberg+witten:1999} (see also
\cite{cornalba+schiappa:2002,matsuo:2002+2001}). The physics of this
case, corresponding to a flat brane embedded in a flat background
space, is well understood \cite{schomerus:1999}. When $\omega$ is
not constant, but it satisfies $\ud\omega=0$, the target space
becomes a Poisson manifold and thus the Kontsevich prescription
\cite{Kontsevich} can be used to define the associative product. In
the most general case, $\ud\omega\neq0$, it has been established
that the extension of the Kontsevich as well as the Moyal products
become nonassociative \cite{cornalba+schiappa:2002}.

Defining gauge theory on nonassociative manifolds is not
straightforward. Recently, there have been attempts to restore the
associativity of the star-product, when the ordinary derivative is
replaced by the covariant derivative. In
\cite{lizzi+mangano+miele+peloso:2002} (see also
\cite{harikumar+rivelles:2006} for details and example) was
considered a curved background and a $\theta^{\mu\nu}$ which is a
covariantly constant antisymmetric tensor, $D_\mu
\theta^{\nu\rho}=0$. However, the appearance of the commutators
$[D_\mu,D_\nu]$ in the star-product, which vanish only on scalar
functions, spoils the associativity. By considering a background
space endowed with a Friedmann-Robertson-Walker metric, it was found
\cite{lizzi+mangano+miele+peloso:2002} that $\theta^{\mu\nu}$ can be
chosen such that the nonassociativity appears at the fourth order in
$\theta$, while the noncommutative effects are already present
starting with $\theta^2$. No extension to a ``complete'' associative
star-product has, however, been obtained.

In \cite{matsuo:2002+2001} it was suggested that such a
nonassocitivity ``anomaly'' can be removed by including the
Chan-Paton factors to define the associative star-product, starting
from the axioms of the rational conformal field theory. It is argued
that by using the vacuum string field theory, one may push most of
the D-branes in a so-called ``closed string vacuum''. In this case
the associativity is restored, i.e. the Chan-Paton factors modify
the originally nonassociative algebra to an associative one. An
infinite number of D-branes are, however, needed for this
modification.

In the gauge theory of the twisted Poincar\'e algebra proposed in
our work, the twist condition \eqref{eq:first-twist-condition} is
not satisfied. This means that the algebra of the twist symmetry
does not close, a property which also implies the nonassociativity
of the star-product.

We believe, however, that in formulating the gauge theory of
noncommutative gravity, the requirement of general coordinate
transformations with respect to the whole Lorentz group should be
relaxed and replaced by the requirement of general coordinate
transformations only under the residual symmetry of the
noncommutative field theories as argued in
\cite{chaichian+nishijima+salminen+tureanu:2008}. This approach will
be pursued in a forthcoming communication \cite{work-in-progress}.
 In a quite different context, the description of
Nature at the Planck scale is suggested to be given by a nonlocal
translationally invariant theory, the so-called "Very Special
Relativity", with a symmetry under a subgroup of the Lorentz group
\cite{VSR}, while at low-energy scale the Poincar\'e invariance
would be operating. A realization of such a symmetry, for the Planck
scale part, has been recently given \cite{NCVSR} on the
noncommutative space-time with light-like noncommutativity. A gauge
theory of the latter symmetry can be performed as mentioned above
\cite{work-in-progress}.


\subsection*{Acknowledgements}

We are indebted to Kazuhiko Nishijima, Ruibin Zhang and Xiao Zhang
for discussions during different stages of this work. A. T.
acknowledges the project no. 121720 of the Academy of Finland. G. Z.
acknowledges the support by the CNCSIS-UEFISCSU grant 620 of the
Minister of Education, Research and Youth of Romania.


\end{document}